\documentclass{article}
\usepackage[german, english]{babel}
\usepackage{a4wide}
\usepackage{amsmath}
\usepackage{amssymb}
\usepackage{ifthen}
\usepackage{epsfig}
\usepackage{setspace}
\usepackage{subfigure}
\usepackage{graphicx}

\newcounter{fig}
\newcommand{\lbfig}[1]{\refstepcounter{fig}
\label{#1} }

\newcommand{\Tr}{{\rm Tr}}

\newcommand{\bea}{\begin{eqnarray}}
\newcommand{\eea}{\end{eqnarray}}
\newcommand{\be}{\begin{equation}}
\newcommand{\ee}{\end{equation}}
\def\diag{\mathop{{\rm diag}}\nolimits}

\def\vecalpha{{\pmb{\alpha}}}
\def\vecbeta{{\pmb{\beta}}}

\def\vecsigma{{\pmb{\sigma}}}

\def\bfph1{{\pmb{\phi}}^{(I)}}
\def\bfph2{{\pmb{\phi}^{(II)}}}

\newcommand{\re}[1]{(\ref{#1})}

\begin{document}

\title{$G_2$ monopoles}
\date{~}
\author{
{\large Ya. Shnir}$^{\dagger \star}$
and {\large G. Zhilin}$^{\star}$ \\
\\ $^{\dagger}${\small BLTP, JINR, Dubna, Russia}
\\ $^{\star}${\small Department of Theoretical Physics and Astrophysics}\\
{\small Belarusian State University, Minsk 220004, Belarus}
} \maketitle

\begin{abstract}

We investigate some aspects of Bogomolny-Prasad-Sommerfield
monopole solutions in the Yang-Mills-Higgs theory with exceptional
gauge group $G_2$ spontaneously broken to $U(1)\times U(1)$.
Corresponding homotopy group is $\pi_2\left(G_2/U(1)\times
U(1)\right)$ and similar to the $SU(3)$ theory, the $G_2$
monopoles are classified by two topological charges $(n_1,n_2)$.
In fundamental representation these yield a subset of $SO(7)$
monopole configurations. Through inspection of the structure of
$Alg(G_2)$, we propose an extension of the Nahm construction to the
$(n,1)_{G_2}$ monopoles. For $(1,1)_{G_2}$ monopole the Nahm data are written
explicitly.
\end{abstract}


\section{Introduction}

Classical monopole solutions of spontaneously broken Yang-Mills-Higgs theories have long been the
objects of detailed study\footnote{For a review, see \cite{Manton-Sutcliffe,Weinberg:2006rq,Shnir}}.
These topologically nontrivial field configurations
may exist in gauge theories for an arbitrary semisimple compact Lie
group \cite{Schwarz:1976xu,Leznov:1979kh}. The simplest example is the 't Hooft-Polyakov
monopole in the $SU(2)$ theory \cite{'tHooft:1974qc,Polyakov:1974ek}.
In the Bogomol'nyi-Prasad-Sommerfield (BPS) limit
\cite{Bogomolny:1975de,Prasad:1975kr} the potential of the scalar field is vanishing and
the monopole solution is given by the first order equation which is integrable. Furthermore,
the Bogomolny equation can be treated as dimensionally reduced self-duality equation and there is
a duality between the monopole solutions of the Bogomolny equation and the matrix valued Nahm data
\cite{Nahm:1979yw}. The Nahm's construction is a very powerful tool for constructing various multimonopoles
in different models \cite{Hurtubise:1989qy,Houghton:1999qu,Irwin:1997ew,Weinberg:1998hn,Lee:1997ny,Houghton:2002bz},
it also has a very interesting realization in the context of construction of D-branes \cite{Diaconescu:1996rk}.\\

Nahm's construction can be generalized for all classical groups, as $SU(N)$ \cite{Weinberg:1998hn}, symplectic and
orthogonal groups \cite{Lee:1997ny,Houghton:2002bz,Lu:1998br}. Here we will concentrate on the case of
the smallest simply connected compact exceptional group
with a trivial center $G_2$. Topologically non-trivial boundary conditions of the scalar field yield nontrivial second
homotopy group of the vacuum where the symmetry is broken to a residue group $H$, this there are monopole solutions of the
$G_2$ Yang-Mills-Higgs theory.\\

Gauge theories with symmetry group $G_2$ have attracted much
attention recently
\cite{Wellegehausen:2009rq,Wellegehausen:2010ai,Landsteiner,Cossu:2007dk,Ilgenfritz:2012wg,Gunaydin:1995ku,Poppitz:2012nz}.
One of the reasons is that such a theory is similar to usual
$SU(3)$ gluodynamics, thus it is useful to investigate how the
center symmetry is relevant for deconfinement phase transition in
the lattice $G_2$ gluodynamics
\cite{Wellegehausen:2009rq,Wellegehausen:2010ai,Cossu:2007dk,Ilgenfritz:2012wg}.
Recently, it was shown that in supersymmetric Yang-Mills theory,
confinement-deconfinement transition does not break the
symmetry of the $G_2$ ground state although the expectation value of the Wilson line exhibits a discontinuity
\cite{Poppitz:2012nz}.

On the other hand, the gauge group $G_2$ is the automorphism group
of the division algebra of octonions. This property allows to
construct octonionic instanton solution to the seven-dimensional
$G_2$ Yang-Mills theory \cite{Gunaydin:1995ku}. Also the massless
monopole states in the $N=2$ supersymmetric  Yang-Mills theory
with symmetry group $G_2$ were considered recently
\cite{Landsteiner}.

Note that coupling of the gauge sector to the Higgs field in the seven-dimensional
fundamental representation of $G_2$ may break this symmetry to $SU(3)$, however in this case
some of fundamental monopoles, i.e. the monopoles associated
with simple roots of the gauge group $G_2$, become massless (see e.g. \cite{Weinberg:2006rq}).
In this paper we will mainly consider another, more simple situation, when the gauge
symmetry is broken maximally by an adjoint Higgs mechanism to $U(1) \times U(1)$. In this case
the monopoles have two topological charges with respect to either of the unbroken Abelian groups $U(1)$, thus the monopoles
can be labeled by two integers $(n_1,n_2)$. \ \

The organization of the paper is as follows. Section II is a review of the basic properties of
the first exceptional group $G_2$, there we also review the Nahm's formalism. Section III
contains our results of construction of the $(1,1)_{G_2}$ monopoles. In
Section IV we conclude with some additional remarks. In additional appendices we summarize the relevant information
about the $\mathfrak{g}_2$ algebra and its representation.

\section{Exceptional group $G_2$ and the Nahm construction}

We start with some introductory remarks about the Lie group
$G_2$. It is the smallest of the five exceptional simple Lie
groups with trivial central element. Mathematically it can be thought as the group of automorphisms of the octonions
or as a subgroup of the real orthogonal group $SO(7)$ which leaves one element of the 8-dimensional
real spinor representation invariant.
It is one of three simple Lie groups of rank two: $SU(3)$, $O(5)$ and $G_2$. The
fundamental representation of $G_2$ is 7-dimensional, the number of generators of the corresponding algebra is
14 (we refer to the appendix A for details). Thus, the Cartan subgroup contains two commuting
generators $H_1,H_2$. The roots and coroots of the $G_2$ are shown in Fig.~\ref{f-1}

\begin{figure}[hbt]
\lbfig{f-1}
\begin{center}
\includegraphics[height=.27\textheight, angle =0]{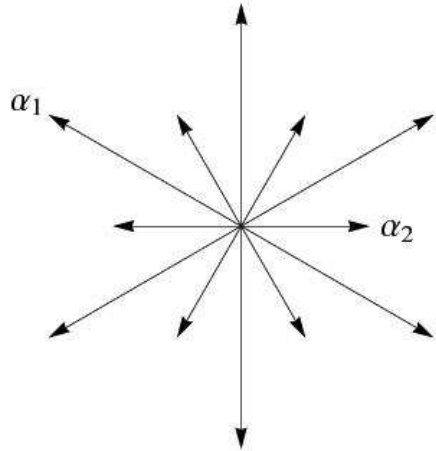}
\end{center}
\caption{\small Root diagram of $G_2$ theory.}
\end{figure}

Explicitly, we can take the elements of the Cartan subalgebra $\mathbf{H}$
\be
\label{Cartan}
H_1 = ~ \frac{1}{4} {\rm diag} (-1,1,-2,0,2,-1,1),\quad H_2 = ~ \frac{1}{4\sqrt{3}} {\rm diag}(0,-1,1,0,-1,1,0);
\ee
so that the Killing form $\mathcal{K} (H_i, H_j) = \frac{1}{2} \delta_{i j}$.

Thereafter we consider the Yang-Mills-Higgs theory in the BPS limit. Then the monopoles are solutions of the first order
Bogomol'nyi equation
\be
\label{BPS}
D_k\Phi = B_k
\ee
The asymptotic value of the Higgs field  along the positive direction of the third axis
lies in the Cartan subalgebra: $\Phi_{\infty} = \mathbf{h} \cdot \mathbf{H }$. If the $G_2$
symmetry is maximally broken to $U(1)\times U(1)$, all roots have non vanishing inner product with
vector $\mathbf{h}$ and, since $\pi_1(G_2)=0$, the monopole solutions
are classified according to the homotopy group
$\pi_2(G_2/U(1)\times U(1)) = \pi_1(U(1)\times U(1)) = \mathbb{Z} \times\mathbb{Z}$.
Recall that the magnetic field of the monopole configuration asymptotically also lies in the
Cartan subalgebra
\be
B_k =  \mathbf{g} \cdot \mathbf{H } \frac{r_k}{4 \pi r^3} \, .
\ee
Therefore the quantized magnetic charge is
\be
\mathbf{g} = \frac{4 \pi}{e} \left(n_1 \vecalpha_1^* + n_2 \vecalpha_2^* \right)
\ee
where two integers $n_1,n_2$ are topological charges of the monopoles given by
embedding along the corresponding simple roots,
there are two distinct charge one fundamental monopoles which correspond to embeddings
along the roots ${\vecalpha}_1$ and ${\vecalpha}_2$, they are (1,0) and (0,1),
respectively. Thus, any $(n_1,n_2)$ $G_2$ monopole can be viewed as a collection of
$n_1$ individual $\vecalpha_1$
fundamental monopoles and $n_2$ $\vecalpha_2$ fundamental monopoles.

Then, making use of an explicit 7-dim representation of $\mathfrak{g}_2$, the asymptotic
of the Higgs field is of the form
\be
\label{Higgsasymptotic}
\begin{split}
\Phi &= ~{\rm diag}(-s_1-s_2, -s_2,-s_1,0,s_1,s_2,s_1+s_2)\\
&- \frac{1}{2er}
~{\rm diag}(-n_2,-n_1+n_2,n_1-2n_2,0,-n_1+2n_2,n_1-n_2,n_2) + O(r^{-1})
\end{split}
\ee
where $s_2 > s_1 >0$ to follow the conventional ordering. The mass of the corresponding
$(n_1,n_2)$ configuration is given by
\be
\label{mass}
M = \frac{4\pi}{e} \left[n_1 \mathbf{h} \cdot \vecalpha_1^* + n_2 {\mathbf{h}} \cdot {{\pmb{\alpha}}_2^*} \right] =
\frac{4\pi}{e} \left[8n_1(s_2-s_1) + 24 n_2 s_1 \right] \, .
\ee

Let us briefly discuss the special case of non-maximal symmetry breaking.
Clearly, there are two situations when one of the $G_2$ monopoles becomes massless, $s_1=s_2$ and $s_1=0$. The first case
corresponds to the situation when the vector of the Higgs field is orthogonal to the long root $\vecalpha_1$ and the
symmetry of broken to $SU(2)\times U(1)$. In the second case the Higgs field is orthogonal to the short root $\vecalpha_2$
and the symmetry is broken to $U(1)\times SU(2)$. The  total magnetic charge of these configurations is Abelian
when the configuration remains
invariant with respect to the transformations from the unbroken subgroup, such configurations are $([3n],2n)$ and
$(2n,[n])$, where the square brackets denote the holomorphic charge which counts the number of
massless monopoles \cite{Lee:1996vz}.

The Nahm construction can be considered as a duality between the
Bogomolny equation \re{BPS} in $\mathbb{R}^3$ and solutions of the Nahm equation in 1-dim space
\be
\label{Hahm}
\frac{d T_i}{ds} = \frac12 \varepsilon_{ijk} [T_j,T_k] \, ,
\ee
where the Nahm data $T_k(s)$ are matrix-valued functions
of a variable $s$ over the finite interval given by
the eigenvalues of the Higgs field on the spacial boundary. The first step of the Hahm construction
is to find a solution of the linear differential
equation \re{Hahm} which must satisfy certain boundary conditions imposed on the
endpoints of the interval of values of variable $s$. The second step is to solve the
construction equation\footnote{Here we consider the $SU(N)$ model.}
on the eigenfunctions $\omega({\bf r},s)$ of the linear operator which includes the Nahm data
\be
\label{constructor}
\left[-\mathbb{I}
_{2k}\frac{d}{ds}+\left(r_i \mathbb{I}
_k  -T_i^{(k)} \right) \otimes \sigma_i\right] \omega({\bf r},s) + (v^{(k)})^\dagger S^{(k)}(\bf r)=0 \, .
\ee
Finally, the
normalizible eigenfunctions allows us to recover the spacetime fields of the BPS monopole as
\be
\label{fields}
\Phi_{nm}=\int\limits_{s_1}^{s_2}\!\! ds~s ~\omega^\dagger_n(s,{\bf r})
\omega_m(s,{\bf r});\qquad A^k_{nm} = -i
\int\limits_{s_1}^{s_2}\!\! ds~~\omega^\dagger_n(s,{\bf r})
\partial^k\omega_m(s,{\bf r})
\ee
where $s_1,s_2$ are the endpoints of the interval of values of variable $s$.

This kind of duality was investigated in many papers,
for a review see \cite{Weinberg:2006rq}, especially in the case of the gauge group $SU(2)$. In such a case
it is possible to prove the isometry between the hyperk\"aler metrics of the
moduli spaces of Nahm data and BPS monopoles. The conjecture about general
equivalence of the metric on the moduli space of
the Hahm data and the metric on the monopole moduli space was used, for example to calculate the
metric on the moduli space of $(2,1)$ $SU(3)$ monopoles \cite{Houghton:1999qu}.

The Nahm approach can be generalized to all classical groups \cite{Hurtubise:1989qy,Lee:1996vz}.
The asymptotic Higgs field of the $SU(N)$ monopoles has $N$ eigenvalues $s_p$, $p=1,2\dots N$ where the usual
ordering is imposed: $s_1 \le s_2 \le \dots \le s_N$. Thus, if the symmetry is broken to maximal torus,
there are $N-1$ fundamental monopoles and the dimension of the corresponding moduli space is $4(N-1)$.
The Nahm data are defined over the interval $s \in [s_1,s_N]$,
this range is subdivided into 6 subintervals $[s_p,s_{p+1}]$ on each of them the Nahm
matrices $T_k(s)$ of dimension $n_p\times n_p$ satisfy the equation
\re{Hahm} \cite{Weinberg:1998hn}. Thus, each
of these subintervals corresponds to a different fundamental monopole, the length of the
subinterval defines its mass and the dimension
of the matrices $T_k(s)$ yields the number of monopoles of that type.

The  boundary conditions on the endpoint of the subintervals are
\begin{enumerate}
\item{
$n_p > n_{p+1}$: $T^{(p+1)}$, should have a well defined limit at $s_{p+1}$, and
\begin{equation}
T^{(p)} = \begin{pmatrix}
T^{(p+1)}(s_{p+1}) + O(s-s_{p+1}) & O\left[ (s-s_{p+1})^{(n_p-n_{p+1}-1)/2} \right] \\
O\left[ (s-s_{p+1})^{(n_p-n_{p+1}-1)/2} \right] & - \dfrac{L^{(p)}}{s-s_{p+1}} + O(1)
\end{pmatrix}
\end{equation}
near the boundary. Here the $n_p\times n_p$ matrix form an irreducible $n_p$-dim representation of $SU(2)$.}
\item $n_p < n_{p+1}$: the roles of the left and right endpoints of the subintervals are reversed and the residue submatrix $L^{(p)}$ appears
in the left upper corner;
\item $n_p = n_{p+1}$: The Nahm data at the endpoint can be discontinuous, one has to introduce the jumping data,
$n_p \times 2$ sized matrix $a$, and require that  at the junction
\begin{equation}
\left( T^{(p+1)}_j - T^{(p)}_j \right)_{rs} = -\frac{1}{2} a^\dag_{s\alpha} (\sigma_j)_{\alpha \beta} a_{\beta r}.
\end{equation}
Here $\sigma_j$ are the usual Pauli matrices.
\end{enumerate}

\section{Construction of the $G_2$ monopoles}
Apart from simple embedding of the properly rescaled $SU(2)$ monopole in the $2\times 2$ block of the $G_2$ matrices
there is another, less trivial embedding into $G_2$. Indeed, $\mathfrak{g}_2$ algebra possesses $\mathfrak{su}(3)$
subalgebra, it can be decomposed as
\begin{equation}
\mathfrak{g}_2 = \mathfrak{su}(3) \oplus \mathfrak{G},
\end{equation}
with $\mathfrak{G}$ forming a module under adjoint action
of $\mathfrak{su}(3)$, $[\mathfrak{su}(3),\mathfrak{G}] = \mathfrak{G}$.

This observation leads to a curious consequence regarding zero modes of the $SU(3)$ embedded
monopole configuration. Indeed, let us consider the
corresponding linearised Bogomol'nyi equation for monopole zero modes
\begin{equation}
\mathcal{D} \delta A = 0.
\end{equation}
Since $\mathcal{D}$ is $\mathfrak{su}(3)$-valued, these modes clearly separate into
purely $\mathfrak{su}(3)$ valued modes and purely $\mathfrak{G}$ ones. The former
are just zero modes of the embedded $SU(3)$ monopole while
the latter appear since $G_2$ is larger than $SU(3)$.
However we can see that the norm of the Higgs field is not affected by excitation of
the $\mathfrak{G}$-valued zero modes:
\begin{equation}
\delta \frac{1}{2}\Tr \Phi^2 = \Tr \Phi \delta \Phi,
\end{equation}
By Ward's formula for energy density of the BPS monopoles \cite{Ward1981},
the excitation of these modes do not change the energy density distribution either. Note that physically these $\mathfrak{G}$-valued zero modes correspond to the decay of certain kind of $SU(3)$ monopoles into a pair of different $G_2$ monopoles.

Let the simple roots of
$\mathfrak{su}(3)$ subalgebra are $\vecbeta_1, \vecbeta_2$.
Their corresponding coroots can be decomposed in coroots of $G_2$ as
\begin{equation}
\label{decompose}
\vecbeta_1^* = \vecalpha_1^*, \hspace{5pt} \vecbeta_2^* = \vecalpha_1^* + \vecalpha_2^*,
\end{equation}
Thus, we can set a correspondence between the monopoles as $(n_1,n_2)_{SU(3)} \rightarrow (n_1+n_2,n_2)_{G_2}$.
In other words, the first
fundamental $SU(3)$ monopole can be viewed as the first fundamental $G_2$ monopole,
and the second as a stack of both
fundamental $G_2$ monopoles. Such identification is somewhat akin to the construction of the $SO, Sp$ monopoles by
restriction of the corresponding $SU(N)$ configurations \cite{Hurtubise:1989qy}, however
the identification of some monopole species in this case happens without reduction of number of species.

This kind of embedding can be used to obtain some non-trivial configurations. For instance, consider the embedding
$(1,1)_{SU(3)} \rightarrow (2,[1])_{G_2}$ (for the second $G_2$ monopole to be massless, original $SU(3)$ monopoles
should be of equal masses). The result is the axially-symmetric subset of the $(2,[1])_{G_2}$ configurations, i.e.
two separated identical monopoles with a cloud of minimal size. We immediately arrive at the conclusion that $(2,[1])_{G_2}$
moduli space interpolates between Taub-NUT (which corresponds to the case of the
non-Abeian cloud of minimal size) and Atiyah-Hitchin (the cloud of infinite size)
geometries. The same result was obtained earlier by another method in \cite{Lee:1997ny} via
identification of certain species of the $SO(8)$ monopoles.

Axially-symmetric $(2,[1])_{SU(3)}$ configurations were studied in
detail in \cite{Dancer1992}. Such configurations can be of two types, the first one
corresponds to the trigonometric axially symmetric Nahm data, it can be considered as the system of
two coincident monopoles surrounded by a non-Abelian cloud
of finite size. The configuration of the second type corresponds to the hyperbolic
axially symmetric Nahm data, then the system is composed of
two separated monopoles with a non-Abelian cloud of minimal size.
By the embedding $(1,1)_{SU(3)}\rightarrow (2,[1])_{G_2}$ we obtain precisely
the latter configuration. Calculating of the energy density profile of the $(1,1)_{SU(3)}$ embedded monopole then
immediately yields the profile of the corresponding axially symmetric
$(2,[1])_{G_2}$ configuration.

Apart this simple embedding, there are different $G_2$ monopoles which can be constructed directly from the Nahm data.
First, let us overview how this formalism can be extended to the
classical groups other than $SU(N)$. Since both $SO(N)$ and $Sp(N)$ groups can be represented by unitary matrices with unit determinant,
the corresponding monopole configurations can be obtained by imposing constraints on a general $SU(N)$ solution. In effect,
these constraints force some species of $SU(N)$ monopoles to merge, reducing the total number of fundamental monopoles.

Our approach to $G_2$ monopoles is essentially the same. Making use of the fundamental 7-dimensional representation we
have established the asymptotic behavior (\ref{Higgsasymptotic}) of $G_2$ monopoles. From Nahm construction point of view,
the leading term of (\ref{Higgsasymptotic}) specifies the intervals on which Nahm matrices are defined. The subleading term tells
us the number of fundamental $SU(7)$ (or $SO(7)$, since $G_2 \subset SO(7)$) monopoles involved. That is, $(n_1,n_2)_{G_2}$ monopoles
lie in the $(n_2,n_1,2 n_2, 2 n_2, n_1, n_2)_{SU(7)}$ sector (more precisely, its $(n_2,n_1,n_2)_{SO(7)}$ subsector). Thus, similar to the
case of orthogonal group, we need to merge further the $(1,0,0)_{SO(7)}$ and $(0,0,1)_{SO(7)}$ monopoles to form the $(0,1)_{G_2}$ monopole.

Note that we can look at the $G_2$ monopoles both from the $SU(7)$ and $SO(7)$ points of view.
The former approach seems to be more natural in the context of Nahm construction, however
the latter approach allows us to deal with less number of the moduli parameters.
Also the $G_2$ is a subgroup of the group $SO(7)$.

Finally, knowing the intervals on which Nahm matrices reside and their dimensions, we
need to place a constraint on the Nahm data directly to merge some monopole species. The
transition from $SU(7)$ to $SO(7)$ is well known, the Nahm matrices should possess a reflection symmetry
\be \label{soconstraint}
T_j(-s) = C(s) T_j^t(s) C^{-1}(s) \, ,
\ee
where the matrix $C(s)$ satisfies $C(-s) = -C^t(s)$. The transition from $SO(7)$ to $G_2$, similar to the construction of the
$SO(N)$ and $Sp(N)$ monopoles via restriction of the $SU(N)$ Nahm data, should relate the
Nahm matrices in the first and the third subintervals (since $(0,1)_{G_2}
\cong (1,0,1)_{SO(7)} \cong (1,0,2,2,0,1)_{SU(7)}$). However, the matrices in these intervals are of different size,
thus, any constraint of the type (\ref{soconstraint}) will not be sufficient.

Some progress can be made if we consider the $(n,1)_{G_2} \cong (1,n,1)_{SO(7)}$ sector. There is only one monopole
of the first and of the third kind, and their coordinates enter the Nahm data explicitly (due to reflection symmetry
only we restrict ourselves to $s\leq 0$):
\begin{align}
T_j(s) = x_j, \hspace{5pt} s\in [-s_1-s_2,-s_2] ,\\
T_j(s) = I_2 y_j + \ldots, \hspace{5pt} s \in [-s_1,0],
\end{align}
where ellipsis denotes the traceless part, determined by the moduli of the $n$ monopoles of the second kind.
Coordinates of the monopoles to be nested are given by $x_j$ and $y_j$, it is natural to conjecture that
the transition from $SO(7)$ to $G_2$ is accomplished by setting $x_j = y_j$. This automatically leaves us with a
correct number of monopole moduli in the Nahm data.

Let us now see how the construction works for the simplest non-trivial case, $(1,1)_{G_2}$. The skyline diagram and the
corresponding Nahm matrices are given in Fig.~\ref{f-3}. For the sake of simplicity the second monopole is placed at the origin.

\begin{figure}[hbt]
\lbfig{f-3}
\begin{center}
\includegraphics[height=.26\textheight, angle =0]{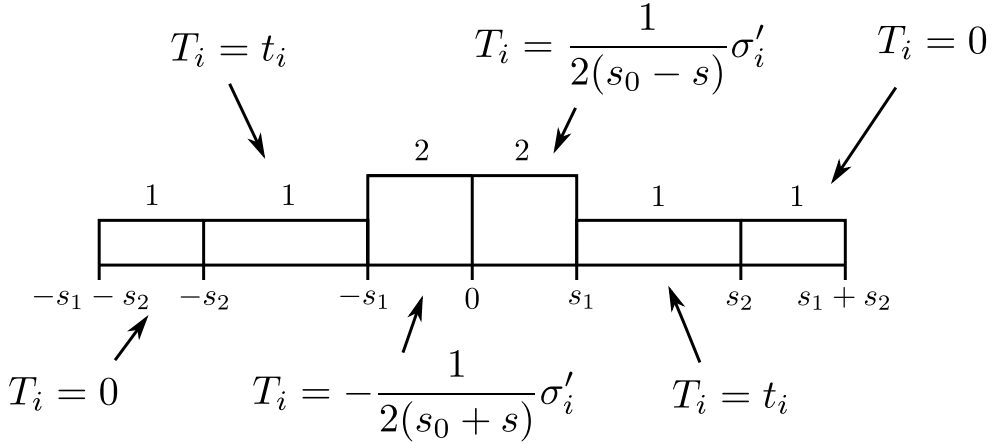}
\end{center}
\caption{\small Skyline diagram of $(1,1)_{G_2}$ monopole and its Nahm data.}
\end{figure}
Here $\sigma'_j = U \sigma_j U^\dag \, (U^\dag = U^{-1})$ are rotated Pauli matrices.
The parameters of the rotation and the value $s_0$ are fixed by the matching condition
across the boundaries of the subintervals
\be
t_i=-\frac{1}{2s_0}(\sigma_i^\prime)_{22}=\frac{1}{2s_0}(\sigma_i^\prime)_{11}
\ee
The Nahm matrices are supplemented by the jumping data

\begin{gather}
s = 0: \hspace{3pt} a_{r\alpha} = \sqrt{\frac{2}{s_0}}U \begin{pmatrix}
0 & -1 \\
1 & 0
\end{pmatrix}, \nonumber \\
s = - s_2: \hspace{3pt} a_\alpha = \sqrt{2 |t_i|} \begin{pmatrix}
\sin \theta/2 \hspace{1pt} e^{-i\varphi/2} \\
-\cos \theta/2 \hspace{1pt} e^{i\varphi/2}
\end{pmatrix},\\
s = + s_2: \hspace{3pt} a_\alpha = \sqrt{2 |t_i|} \begin{pmatrix}
\cos \theta/2 \hspace{1pt} e^{-i \varphi/2} \\
\sin \theta/2 \hspace{1pt} e^{i \varphi/2}
\end{pmatrix}, \nonumber
\end{gather}
where $\theta$ and $\varphi$ specify direction of $t_i$.

It is a trivial matter to carry out the construction in the $t_i = 0$ case. The two fundamental $G_2$ monopoles now
coincide, they are
spherically symmetric. This case corresponds to the $SU(3)$ composite monopole embedded along the
root $\vecbeta_3 = \vecbeta_1 + \vecbeta_2$.
Then the complete orthonormal set of construction equation solutions can be taken to be
\be
\begin{split}
\omega_1&=\sqrt{\frac{r}{ \sinh v r}} \exp(s \sigma_i \cdot r_i - \frac{r s_2}{2})\eta_-^{down};
\qquad \omega_2=0,~~ S(-s_2)=1;\\
\omega_3&=\sqrt{\frac{r}{ \sinh v r}} \exp(s \sigma_i
\cdot r_i + \frac{r s_2}{2})\eta_-^{up};
\qquad \omega_4=0,~~ S(0)=1;\\
\omega_5&=\sqrt{\frac{r}{ \sinh v r}} \exp(s \sigma_i
\cdot r_i + \frac{r s_2}{2})\eta_+^{down};
\qquad \omega_6=0,~~ S(s_2)=1;\\
\omega_7&=\sqrt{\frac{r}{ \sinh v r}} \exp(s \sigma_i
\cdot r_i - \frac{r s_2}{2})\eta_+^{up} \, .
\end{split}
\ee
where $\eta_{\pm}^{up/down}$ are the usual eigenvectors of $\vecsigma \cdot \bf r$.
These solutions give rise to the Higgs field of the $G_2$ monopole
\begin{align}
\Phi =& s_2 \hspace{2pt}\diag (-\frac{1}{2},-1,\frac{1}{2},0,-\frac{1}{2},1,\frac{1}{2}) \nonumber \\
& + \frac{1}{2} \left[ (2s_1 + s_2)\coth (2s_1 + s_2)r - \frac{1}{r}  \right] \diag (-1,0,-1,0,1,0,1).
\end{align}
One can readily recognize the Higgs profile of a spherically symmetric monopole in the string gauge.
We obtain the fields of an embedded $SU(2)$ monopole, just as expected.

For non-zero separation the construction equation can be solved analytically, however picking an orthonormal
basis of its solutions is a technically difficult task.

In this simple case we can check the correctness of the construction indirectly.
The $(1,1)_{G_2}$ solution is obtained by placing a constraint on a generic $(1,1,1)_{SO(7)}$ monopoles.
Both configurations contain no more than one monopole of each kind. Thus, the corresponding asymptotic metrics, which
include monopole coordinates $\mathbf{x}_i$ and phases $\xi_i$, turn out to be exact. This conclusion can be proven rigorously
for two monopoles, since hyperk\"ahler structure and asymptotic interaction completely determines the metric on the moduli space.
On the other hand, the constraint we imposed, selects a submanifold
in $(1,1,1)_{SO(7)}$ moduli space (by setting $\mathbf{x}_1 = \mathbf{x}_3$), and hence gives us
an expression for the metric of $(1,1)_{G_2}$.
Direct computation confirms that the metric obtained by such identification is the correct one.

\section{Conclusions}

The main purpose of this work was to present the application of the Nahm construction to the case of the
BPS monopoles in the Yang-Mills-Higgs theory with exceptional gauge group $G_2$ spontaneously broken to
$U(1) \times U(1)$. As a particular example we considered the Abelian spherically symmetric
$(1,1)_{G_2}$ monopole. We have shown that the $G_2$ monopoles can be constructed by identification of certain set of
$SU(7)$ (or $SO(7)$) fundamental monopoles, in particular the
first  $G_2$ fundamental monopole $(1,0)$ represents a set of two nested $SU(7)$ monopoles
location and orientation of those coincide, while the second
$G_2$ fundamental monopole $(0,1)$ represents another collection of six aligned and nested $SU(7)$ monopoles.

Perhaps the most interesting feature of the Nahm construction is
its realization in the terms of Dirichlet branes. It was pointed
out by Diakonesky \cite{Diaconescu:1996rk} that there is
one-to-one correspondence between the $SU(N)$  monopole embedded
along the simple roots as
$\mathbf{g} = \frac{4 \pi}{e}
\sum\limits_{i} n_i \vecalpha_i^*$ and the 1-branes stretching
between the three-branes separated in a transverse direction. This
sort of duality has been explicitly realized in ${\cal N}=4$
$SU(N)$ super Yang-Mills theory \cite{Hanany:1996ie}.  From that
point of view, the construction of the Nahm data for $G_2$
monopoles corresponds to the configuration of the D-branes some of
which must be identified according to the restrictions
\re{decompose} \cite{Selivanov:2003xg}.

There are various possible applications of the $G_2$ monopole solutions discussed in this work.
An interesting task would be to study the contribution of these configurations in the
confinement-deconfinement phase transitions. Note that this transition in the supersymmetric
$G_2$ Yang-Mills theory recently was discussed in \cite{Poppitz:2012nz}.
In particular, it was shown that deconfinement transition does not break the
symmetry of the $G_2$ ground state although the expectation value of the Wilson line exhibits a discontinuity.

Certainly, this is a first step towards comprehensive study of the monopoles in the gauge models with exceptional groups.
As a direction for future work, it would be interesting to study in more details the moduli space of $G_2$ monopoles,
considering in particular, various cases of non-maximal symmetry breaking. It would allow us
to better understand the role of the corresponding massless $G_2$ monopoles (non-Abelian clouds). Explicit construction of the
$(n_1,n_2)_{G_2}$ moduli space metric, which determines the low-energy of the monopoles, remains our first goal.
We hope to  report elsewhere on these problems.

\section*{Acknowledgements}
We thank Sasha Gorsky, Derek Harland, Evgeny Ivanov, Olaf Lechtenfeld, Nick Manton, Andrey Smilga and Paul Sutcliffe
for many useful discussions and valuable comments.
This work is supported in part by the A.~von Humboldt Foundation in
the framework of the Institutes linkage Programm
and by the JINR Heisenberg-Landau Program (Y.S.). We are
grateful to the Institute of Physics at the Carl von Ossietzky University
Oldenburg for hospitality.

\pagebreak
\section*{Appendix A: $\mathfrak{g}_2$ algebra and its representation}
\resizebox{\linewidth}{!}{
\begin{tabular}{c||c|c|c|c|c|c|c|c||c|c|c|c|c|c}
 &$h_1$&$h_2$&$g_{1,-2}$&$g_{2,-3}$&$g_{3,-1}$&$g_{2,-1}$&$g_{3,-2}$&$g_{1,-3}$&$g_1$&$g_2$&$g_3$&$g_{-1}$&$g_{-2}$&$g_{-3}$\\
\hline
\hline
$h_1$ & $0$ & $0$ & $2g_{1,-2}$ & $-g_{2,-3}$ & $-g_{3,-1}$ & $-2g_{2,-1}$ & $g_{3,-2}$ & $g_{1,-3}$ & $-g_1$ & $+g_2$ & $0$ & $g_{-1}$ & $-g_{-2}$ & $0$ \\
\hline
$h_2$ &  & $0$ & $-g_{1,-2}$ & $2g_{2,-3}$ & $-g_{3,-1}$ & $g_{2,-1}$ & $-2g_{3,-2}$ & $g_{1,-3}$ & $0$ & $-g_2$ & $g_3$ & $0$ & $g_{-2}$ & $-g_{-3}$ \\
\hline
$g_{1,-2}$ &  &  & $0$ & $g_{1,-3}$ & $-g_{3,-2}$ & $h_1$ & $0$ & $0$ & $-g_2$ & $0$ & $0$ & $0$ & $g_{-1}$ & $0$ \\
\hline
$g_{2,-3}$ &  &  &  & $0$ & $g_{2,-1}$ & $0$ & $h_2$ & $0$ & $0$ & $-g_3$ & $0$ & $0$ & $0$ & $g_{-2}$ \\
\hline
$g_{3,-1}$ &  &  &  &  & $0$ & $0$ & $0$ & $-h_1-h_2$ & $0$ & $0$ & $-g_1$ & $g_{-3}$ & $0$ & $0$ \\
\hline
$g_{2,-1}$ &  &  &  &  &  & $0$ & $g_{3,-1}$ & $-g_{2,-3}$ & $0$ & $-g_1$ & $0$ & $g_{-2}$ & $0$ & $0$ \\
\hline
$g_{3,-2}$ &  &  &  &  &  &  & $0$ & $-g_{1,-2}$ & $0$ & $0$ & $-g_2$ & $0$ & $g_{-3}$ & $0$ \\
\hline
$g_{1,-3}$ &  &  &  &  &  &  &  & $0$ & $-g_3$ & $0$ & $0$ & $0$ & $0$ & $-g_{-1}$ \\
\hline
\hline
$g_1$ &  &  &  &  &  &  &  &  & $0$ & $2g_{-3}$ & $-2g_{-2}$ & $2h_1+h_2$ & $3g_{2,-1}$ & $3g_{3,-1}$ \\
\hline
$g_2$ &  &  &  &  &  &  &  &  &  & $0$ & $2g_{-1}$ & $3g_{1,-2}$ & $-h_1+h_2$ & $3g_{3,-2}$ \\
\hline
$g_3$ &  &  &  &  &  &  &  &  &  &  & $0$ & $3g_{1,-3}$ & $3g_{2,-3}$ & $-h_1-2h_2$ \\
\hline
$g_{-1}$ &  &  &  &  &  &  &  &  &  &  &  & $0$ & $2g_3$ & $-2g_2$ \\
\hline
$g_{-2}$ &  &  &  &  &  &  &  &  &  &  &  &  & $0$ & $2g_1$ \\
\hline
$g_{-1}$ &  &  &  &  &  &  &  &  &  &  &  &  &  & $0$ \\
\end{tabular}
}
\vspace{\baselineskip}

Our choice of simple roots is $\vecalpha_1  = g_{1,-2}$ (long root) and $\vecalpha_2 = g_{-2}$ (short root);
$h_{\vecalpha_1^*} = h_1$, $h_{\vecalpha_2^* } = h_2 - h_1$.
The representation is chosen in such a way that the elements of the Cartan
subgroup $h$ with $\vecalpha_{1,2}(h) \geq 0$ have properly ordered eigenvalues.

\begin{align*}
h_1 &= -e_{22} + e_{33} - e_{55} + e_{66}, \\
h_2 &= -e_{11} - e_{33} + e_{55} + e_{77}, \\
g_{1,-2} &= -e_{32} + e_{65}, \\
g_{1,-3} &= e_{61} - e_{72}, \\
g_{2,-3} &= e_{51} - e_{73}, \\
g_1 &= e_{13} - \sqrt{2}e_{24} + \sqrt{2}e_{46} - e_{57}, \\
g_2 &= -e_{12} - \sqrt{2}e_{34} + \sqrt{2}e_{45} + e_{67}, \\
g_3 &= \sqrt{2}e_{41} + e_{52} - e_{63} - \sqrt{2}e_{74}, \\
g_{2,-1} &= (g_{1,-2})^T, \\
g_{3,-1} &= (g_{1,-3})^T, \\
g_{3,-2} &= (g_{2,-3})^T, \\
g_{-1} &= -(g_1)^T, \\
g_{-2} &= -(g_2)^T, \\
g_{-3} &= -(g_3)^T,
\end{align*}
where $e_{nm}$ is $7\times 7$ matrix with the only non-zero element $(e_{nm})_{nm} = 1$.

\pagebreak

\section*{Appendix B: representation of $\mathfrak{su}(3)$ subgroup}
The representation is chosen so that vacuum expectation value of the Higgs field has properly ordered eigenvalues.
\begin{gather*}
h_1 = \begin{pmatrix}
-1 & 0 & 0 \\
0 & 1 & 0 \\
0 & 0 & 0
\end{pmatrix}
h_2 = \begin{pmatrix}
0 & 0 & 0 \\
0 & -1 & 0 \\
0 & 0 & 1
\end{pmatrix} \\
g_{2,-1} = \begin{pmatrix}
0 & \frac{1}{\sqrt{2}} & 0 \\
0 & 0 & 0 \\
0 & 0 & 0
\end{pmatrix}
g_{1,-2}=\begin{pmatrix}
0 & 0 & 0 \\
\frac{1}{\sqrt{2}} & 0 & 0 \\
0 & 0 & 0
\end{pmatrix} \\
g_{3,-2} = \begin{pmatrix}
0 & 0 & 0 \\
0 & 0 & \frac{1}{\sqrt{2}} \\
0 & 0 & 0
\end{pmatrix}
g_{2,-3} = \begin{pmatrix}
0 & 0 & 0 \\
0 & 0 & 0 \\
0 & \frac{1}{\sqrt{2}} & 0
\end{pmatrix} \\
g_{3,-1} = \begin{pmatrix}
0 & 0 & \frac{1}{\sqrt{2}} \\
0 & 0 & 0 \\
0 & 0 & 0
\end{pmatrix}
g_{1,-3} = \begin{pmatrix}
0 & 0 & 0 \\
0 & 0 & 0 \\
\frac{1}{\sqrt{2}} & 0 & 0
\end{pmatrix}
\end{gather*}


\begin{thebibliography}{99}
\bibitem{Manton-Sutcliffe}
N.~S.~Manton and P.~Sutcliffe, ``Topological solitons'',
Cambridge, UK: Univ. Press. (2004) 493 p
\bibitem{Weinberg:2006rq}
E.~J.~Weinberg and P.~Yi,
Phys.\ Rept.\  {\bf 438} (2007) 65
\bibitem{Shnir}
Y.~M.~Shnir,
``Magnetic monopoles''
Berlin, Germany: Springer (2005) 532 p
\bibitem{Schwarz:1976xu}
A.~S.~Schwarz,
Nucl.\ Phys.\ B {\bf 112} (1976) 358
\bibitem{Leznov:1979kh}
A.~N.~Leznov and M.~V.~Saveliev,
Lett.\ Math.\ Phys.\  {\bf 3} (1979) 207
\bibitem{'tHooft:1974qc}
  G.~'t Hooft,
  Nucl.\ Phys.\ B {\bf 79} (1974) 276
\bibitem{Polyakov:1974ek}
A.~M.~Polyakov,
JETP Lett.\  {\bf 20} (1974) 194
[Pisma Zh.\ Eksp.\ Teor.\ Fiz.\  {\bf 20} (1974) 430]
\bibitem{Bogomolny:1975de}
E.~B.~Bogomolny,
Sov.\ J.\ Nucl.\ Phys.\  {\bf 24} (1976) 449
[Yad.\ Fiz.\  {\bf 24} (1976) 861]
\bibitem{Prasad:1975kr}
M.~K.~Prasad and C.~M.~Sommerfield,
Phys.\ Rev.\ Lett.\  {\bf 35} (1975) 760
\bibitem{Nahm:1979yw}
  W.~Nahm,
  Phys.\ Lett.\ B {\bf 90} (1980) 413;\\
W.~Nahm, in "Monopoles in Quantum Field Theory"
edited by N. Craigie et al.
~World Scientific, Singapore, 1982
\bibitem{Hurtubise:1989qy}
  J.~Hurtubise and M.~K.~Murray,
  Commun.\ Math.\ Phys.\  {\bf 122} (1989) 35
\bibitem{Houghton:1999qu}
  C.~Houghton, P.~W.~Irwin and A.~J.~Mountain,
  JHEP {\bf 9904} (1999) 029
\bibitem{Irwin:1997ew}
P.~Irwin,
Phys.\ Rev.\ D {\bf 56} (1997) 5200
\bibitem{Weinberg:1998hn}
  E.~J.~Weinberg and P.~Yi,
  Phys.\ Rev.\ D {\bf 58}, 046001 (1998)
\bibitem{Lee:1997ny}
  K.~M.~Lee and C.~Lu,
  Phys.\ Rev.\ D {\bf 57} (1998) 5260
\bibitem{Houghton:2002bz}
  C.~J.~Houghton and E.~J.~Weinberg,
  Phys.\ Rev.\ D {\bf 66} (2002) 125002
\bibitem{Diaconescu:1996rk}
  D.~E.~Diaconescu,
  Nucl.\ Phys.\ B {\bf 503} (1997) 220
\bibitem{Lu:1998br}
  C.~H.~Lu,
  Phys.\ Rev.\ D {\bf 58} (1998) 125010
\bibitem{Wellegehausen:2009rq}
  B.~H.~Wellegehausen, A.~Wipf and C.~Wozar,
  Phys.\ Rev.\ D {\bf 80} (2009) 065028
\bibitem{Wellegehausen:2010ai}
B.~H.~Wellegehausen, A.~Wipf and C.~Wozar,
  Phys.\ Rev.\ D {\bf 83} (2011) 016001
\bibitem{Landsteiner}
K.~Landsteiner, J.M.~Pierre and S.B.~Giddings.
Phys.\ Rev.\ D {\bf 55} (1997) 2367
\bibitem{Cossu:2007dk}
  G.~Cossu, M.~D'Elia, A.~Di Giacomo, B.~Lucini and C.~Pica,
  JHEP {\bf 0710} (2007) 100
\bibitem{Ilgenfritz:2012wg}
  E.~M.~Ilgenfritz and A.~Maas,
  Phys.\ Rev.\ D {\bf 86} (2012) 114508
\bibitem{Gunaydin:1995ku}
M.~Gunaydin and H.~Nicolai,
Phys.\ Lett.\ B {\bf 351} (1995) 169
[Addendum-ibid.\ B {\bf 376} (1996) 329]
\bibitem{Lee:1996vz}
  K.~M.~Lee, E.~J.~Weinberg and P.~Yi,
  Phys.\ Rev.\ D {\bf 54} (1996) 6351
\bibitem{Ward1981}R.S.~Ward, Commun.\ Math.\ Phys.\ {\bf 79} (1981) 317
\bibitem{Dancer1992}A.S.~Dancer, Nonlinearity\ {\bf 5} (1992) 1355
\bibitem{Diaconescu:1996rk}
  D.~E.~Diaconescu,
  Nucl.\ Phys.\ B {\bf 503} (1997) 220
\bibitem{Hanany:1996ie}
  A.~Hanany and E.~Witten,
  Nucl.\ Phys.\ B {\bf 492} (1997) 152
\bibitem{Selivanov:2003xg}
  K.~G.~Selivanov and A.~V.~Smilga,
  JHEP {\bf 0312} (2003) 027
\bibitem{Poppitz:2012nz}
  E.~Poppitz, T.~Sch\"afer and M.~\"Unsal,
  JHEP {\bf 1303} (2013) 087
\end{thebibliography}
\end{document}